\documentclass[10pt,journal,compsoc]{IEEEtran}

\usepackage{bibnames}
\usepackage{amssymb}
\usepackage{verbatim}
\usepackage{amsmath}
\usepackage{amsthm}
\usepackage[pdftex]{graphicx}
\usepackage[font=scriptsize]{caption}
\usepackage{subcaption}
 \usepackage[linesnumbered, ruled,vlined]{algorithm2e}
\usepackage{listings}
\usepackage[flushleft]{threeparttable}
\usepackage{multirow}
\usepackage{listings}
\usepackage{cite}
\usepackage{hyperref}
\usepackage{mdframed}
\usepackage{tabularx}
\usepackage{booktabs}
\usepackage{flushend}
\usepackage{mathrsfs}
\usepackage[T1]{fontenc}
\usepackage[switch]{lineno}
\usepackage{enumitem}

\usepackage{balance}

\definecolor{light-gray}{gray}{0.92}  
\newenvironment{gtheorem}%
  {\begin{mdframed}[backgroundcolor=light-gray]\begin{mdtheorem}{name}{label}}%
  {\end{mdtheorem}\end{mdframed}}

\usepackage[table]{xcolor}
\definecolor{ao}{rgb}{0.0, 0.5, 0.0}
\newtheorem{definition}{Definition}[section]

\newcommand{\etal}{\textit{et al.}\space}
\newcommand{\cross}{\textit{cross-project}\xspace}
\newcommand{\dev}{\textit{dev.-process}\xspace}
\newcommand{\tool}{\textsc{CodeJIT}\xspace}
\newcommand{\jit}{just-in-time\space}

\begin{document}

\title{Code-centric Learning-based\\Just-In-Time Vulnerability Detection}

\date{}

\author{Son~Nguyen, Thu-Trang~Nguyen,
        Thanh~Trong~Vu, Thanh-Dat~Do, Kien-Tuan~Ngo,
        and~Hieu~Dinh~Vo 
\IEEEcompsocitemizethanks{
    \IEEEcompsocthanksitem Son~Nguyen, Thu-Trang~Nguyen,
        Thanh~Trong~Vu, Thanh-Dat~Do, Kien-Tuan~Ngo,
        and~Hieu~Dinh~Vo are with Faculty of Information Technology, University of Engineering and Technology, Vietnam National University, Hanoi, Vietnam.
    \IEEEcompsocthanksitem Corresponding author: Hieu~Dinh~Vo. \protect\\
    E-mail: \href{mailto:hieuvd@vnu.edu.vn}{hieuvd@vnu.edu.vn}.
    }
}

\IEEEtitleabstractindextext{%
\begin{abstract}
Attacks against computer systems exploiting software vulnerabilities can cause substantial damage to the cyber-infrastructure of our modern society and economy.
To minimize the consequences, it is vital to detect and fix vulnerabilities as soon as possible. 
Just-in-time vulnerability detection (JIT-VD) discovers vulnerability-prone (``dangerous") commits to prevent them from being merged into source code and causing vulnerabilities. By JIT-VD, the commits' authors, who understand the commits properly, can review these dangerous commits and fix them if necessary while the relevant modifications are still fresh in their minds.
In this paper, we propose \tool, a novel code-centric learning-based approach for just-in-time vulnerability detection. 
The key idea of \tool is that \textit{the meaning of the code changes of a commit is the direct and deciding factor for determining if the commit is dangerous for the code.}
Based on that idea, we design a novel graph-based representation to represent the semantics of code changes in terms of both code structures and program dependencies. 
%
%
A graph neural network model is developed to capture the meaning of the code changes represented by our graph-based representation and learn to discriminate between dangerous and safe commits. 
We conducted experiments to evaluate the JIT-VD performance of \tool on a dataset of 20K+ dangerous and safe commits in 506 real-world projects from 1998 to 2022. Our results show that \tool significantly improves the state-of-the-art JIT-VD methods by up to 66\% in Recall, 136\% in Precision, and 68\% in F1. 
Moreover, \tool correctly classifies nearly 9/10 of dangerous/safe (benign) commits and even detects 69 commits that fix a vulnerability yet produce other issues in source code.

\end{abstract}

}

\maketitle
\IEEEdisplaynontitleabstractindextext

\IEEEpeerreviewmaketitle

\section{Introduction}

Software is a critical element in a broad range of real-world systems. Attacks against computer systems exploiting software vulnerabilities (security defects/bugs), especially in critical systems such as traffic control, aviation coordination, chemical/nuclear industrial operation, and national security systems, can cause substantial damage. 
The latest survey~\cite{survey_3} in 2021 shows that by 2020, the total cost of bugs related to software vulnerabilities in the US (accounting for 33\% of the market share in the whole world technology market) is 1.56 trillion USD. 
This cost accounts for 75\% of the cost caused by poor-quality software.
To minimize the consequences caused by vulnerabilities in software, vulnerabilities should be detected and fixed as soon as possible because late detection and correction can cost up to 200 times as much as early correction or even cause more severe damage~\cite{jit_vul_less_is_more,jit_bug,jit_vul_survey}. 
The most popular and effective solution being applied in practice is to use techniques that scan the entire source code and automatically detect the presence of vulnerabilities before the software is released~\cite{survey_papers,ese_empirical}. 

Most of the existing methods scan all the software components such as files, functions, or lines in source code to detect vulnerabilities~\cite{survey_papers,linevd,linevul,ivdetect,velvet,vuldeeppeaker,mu_vuldeeppeker,devign,vulsniper,mvd,issta_22}. These scanning techniques often report a large number of alarms and multiple of them are false alarms~\cite{kang2022detecting, ngo2021ranking}. After that, the detected suspicious components still need to be manually examined to determine the actual presence of the vulnerability. 
Such task requires selecting an \textit{appropriate developer group} who can thoroughly understand the business logic of the components.
%
%
%
However, this proper developer selection task is also very challenging~\cite{jit_vul_less_is_more,jit_bug,jit_vul_survey}.
In practice, developers usually commit their changes very regularly. In Mozilla Firefox, the developers recently pushed 10K+commits to construct a release from the previous version~\footnote{From \texttt{release\_v109} to \texttt{release\_v110}, more details: \url{https://github.com/mozilla-mobile/fenix}}. 
At the same time, we found that there are some files touched by up to 155 developers in Mozilla Firefox. Thus, with large accumulated sets of commits and authors at the released time, correctly identifying the authors of dangerous code could be non-trivial~\cite{vccfinder, jit_bug,jit_vul_less_is_more,szz}. 
Even if the authors are correctly identified, the developers might still need to spend much time recalling and drilling down to the files/functions. This could significantly slow down the quality assurance process and development cycle.
Thus, although the recommendations based on vulnerability detection at the release time (long-term recommendations) can be useful in some contexts, they have their own drawbacks in practice~\cite{jit_bug,jit_vul_less_is_more,jit_vul_survey}. 

Meanwhile, \textit{\jit}~or \textit{commit-level} recommendations (short-term recommendations) could be preferred because they allow commits' authors to get immediate feedback on their code changes. This feedback could help developers faster improve their code quality since the context of the changes is still fresh in their minds. In addition, the \jit vulnerability detection techniques facilitate not only the authors in responsibly making code changes but also code auditors in reviewing code commits~\cite{jit_vul_less_is_more,ese_empirical}. 

The past few years have witnessed a few successful research to investigate \textit{\jit vulnerability detection}~\cite{vuldigger,vccfinder} and \jit general defect detection~\cite{jitline, jit-fine,lapredict,deepjit}.
%
%
These studies utilize commit messages and expert features such as authors' experiences, commit time, and code changes' complexity (e.g., numbers of added/deleted lines of code) to calculate the metrics to determine if commits are suspicious. 
In fact, these features might have some correlations but not necessarily cause the presence of vulnerabilities at the commit level. For example, a commit whose message contains bug-fixing keywords such as ``\textit{fix}", ``\textit{failures}", or ``\textit{resolves}" could be very likely to be classified as a safe one~\cite{lapredict}. 
Although the commit fixes some bugs, the possibility that the commit creates other bugs in the code should not be eliminated.
Meanwhile, the studies utilizing the expert features often leverage on certain hypotheses~\cite{vccfinder,jitline,jit-fine,are_we_there,lapredict} such as the larger number of added lines of code, the higher possibility the commits contain vulnerabilities~\cite{lapredict}. Consequently, the dangerous commits with a few added lines of code could be misidentified.
%
%
Thus, commit messages and the expert features could correlate but not necessarily result in the presence of vulnerabilities, which are, in fact, brought about by code changes.

In this paper, we introduce \tool, a novel \textit{\textbf{code-centric}} learning-based approach which focuses on capturing the meaning of code changes in known dangerous or safe (benign) commits, rather than utilizing commit messages or the expert features, to detect vulnerabilities at the commit level.
Our idea is that \textit{for a commit, the meaning of the changes in the source code is the direct and deciding factor for assessing the commit's suspiciousness.}
%
%
%
This idea is reasonable because a commit should be considered dangerous if its code changes carry vulnerabilities or interact with the unchanged code to collectively introduce vulnerabilities once the changed code is merged into the source code. 
%
%
%

To implement our code-centric idea for JIT-VD, we propose a novel graph-based code transformation representation (\textbf{C}ode \textbf{T}ransformation \textbf{G}raph - CTG) to capture the semantics of the code changes caused by commits by representing the changed code in relation to the related unchanged code. Particularly, the CTG of a commit expresses the code changes/transformations in both the code structure and dependencies which are necessary to capture the syntax and semantics of the code changes.
Based on the code before and after the commit, the CTG is constructed from the set of nodes which are \textit{added}, \textit{deleted}, and \textit{unchanged} code elements such as statements, expressions, or operators. The CTG's nodes are connected by the edges which are the \textit{added}, \textit{deleted}, and \textit{unchanged} structure and dependency relations between the nodes.
Next,
\tool relies on the code-related aspects of commits, which are represented by their CTGs, to learn discriminating dangerous/safe commits. Particularly, CTGs can be intuitively treated as relational graphs. A Relational Graph Convolutional Network (RGCN) model is developed to capture the knowledge from the CTGs of the prior known safe/dangerous commits and learn to detect dangerous commits. 

We conducted experiments to evaluate the JIT-VD performance of {\tool} on a large dataset of 20K+ dangerous/safe commits in 506 C/C++ projects from 1998 to 2022. We compared {\tool} with the state-of-the-art JIT-VD approaches~\cite{vccfinder, deepjit, jitline, lapredict, jit-fine}. 
Our results show that \tool significantly improves the state-of-the-art JIT-VD techniques by up to 68\% in F1 and 77\% in classification accuracy. Especially, \tool can correctly distinguish nearly 9/10 of dangerous/safe commits. The recall of \tool is up to 66\% better than that of the existing approaches. Moreover, \tool can achieve the precision of 90\%, which is 32--123\% better than the precision of the other approaches. In other words, 9/10 suspicious commits detected by \tool actually are dangerous. Hence, with \tool, practitioners could discover more the commits dangerous for their code and spend less undesirable efforts for the false alarms during the security inspection process.
In brief, this paper makes the following contributions:

\begin{enumerate}
    \item Code Transformation Graph: A novel graph-based representation of code changes to capture the changes in the crucial aspects of code: the code structure and program dependencies. 
    \item {\tool}: A novel code-centric approach for detecting vulnerabilities at the commit level.
    \item An extensive experimental evaluation showing the performance of {\tool} over the state-of-the-art methods for just-in-time vulnerability detection.
    \item A public dataset of 20K+ safe and dangerous commits collected from 506 real-world projects, which can be used as a benchmark for evaluating related work.
\end{enumerate}
The detailed implementation of \tool and dataset can be found at: \textit{\url{https://github.com/ttrangnguyen/CodeJIT}}.

\section{Motivating Example}
\label{sec:example}
We first illustrate the problem of \jit vulnerability detection and explain the motivation for our solution.

\subsection{Observations}

\begin{figure}
     \centering
     \begin{subfigure}[b]{0.5\textwidth}
        \centering
        \includegraphics[width=1\columnwidth]{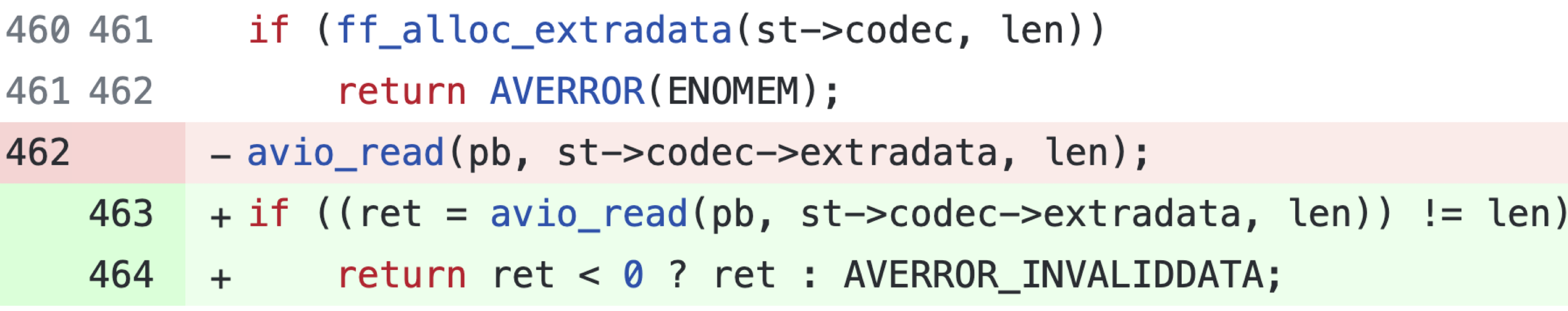}
        \caption{Commit \texttt{2c635fa} fixing a vulnerability yet causing another}
        \label{fig:fixing-vt}
     \end{subfigure}
     \hfill

     \begin{subfigure}[b]{0.5\textwidth}
        \centering
        \includegraphics[width=1\columnwidth]{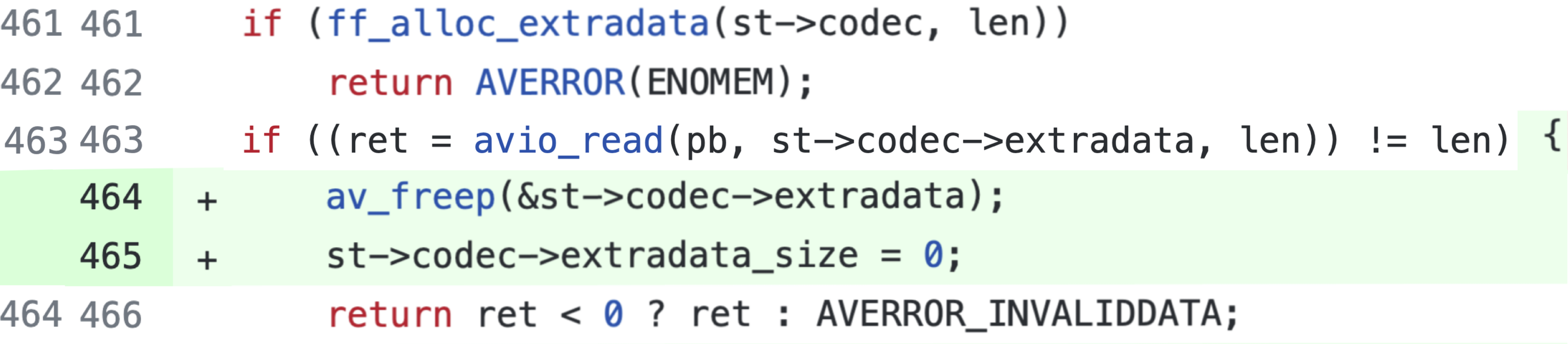}
        \caption{Commit \texttt{ac480cb} fixing the vulnerability in Fig.~\ref{fig:fixing-vt}}
        \label{fig:fixing}
     \end{subfigure}
     
    \caption{Two commits in \textit{FFmpeg} project demonstrating the importance of understanding code changes for JIT-VD.}
\end{figure}

\begin{figure}
     \centering
     \begin{subfigure}[b]{0.5\textwidth}
        \centering
        \includegraphics[width=0.9\columnwidth]{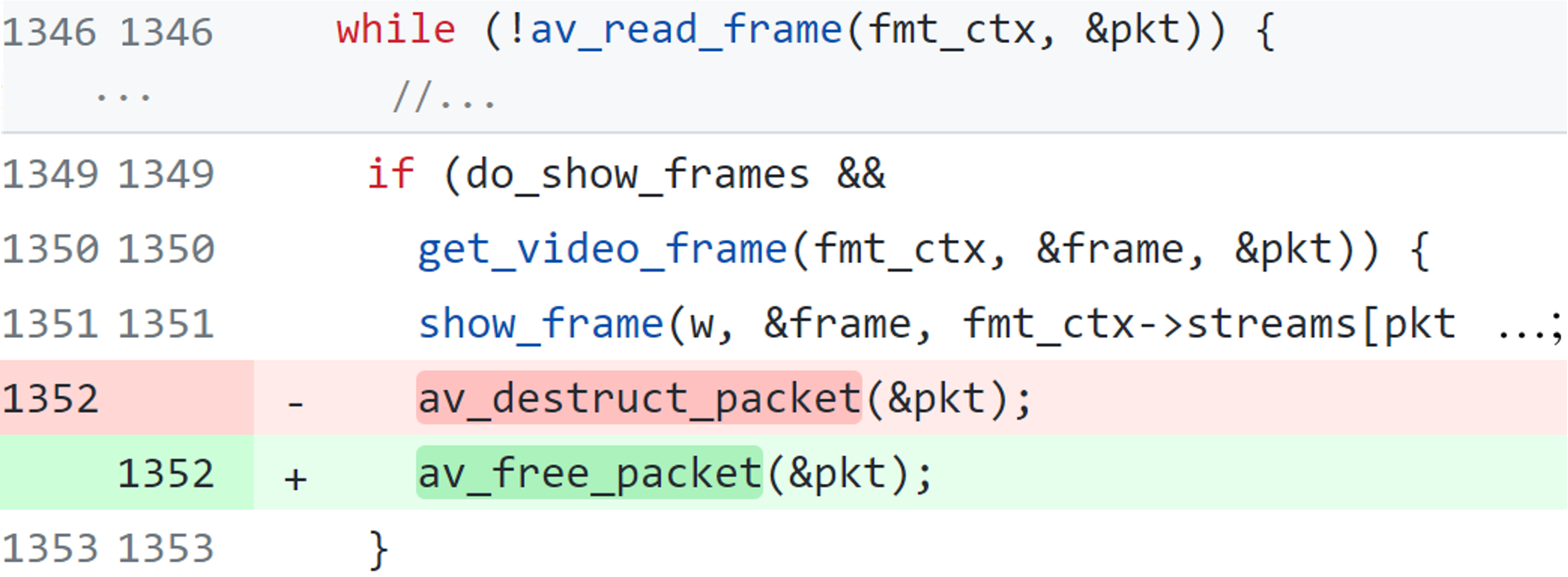}
        \caption{Commit \texttt{7328c2f} causing a vulnerability in \textit{FFmpeg}}
        \label{fig:vt2}
     \end{subfigure}
     \hfill

     \begin{subfigure}[b]{0.5\textwidth}
        \centering
        \includegraphics[width=0.9\columnwidth]{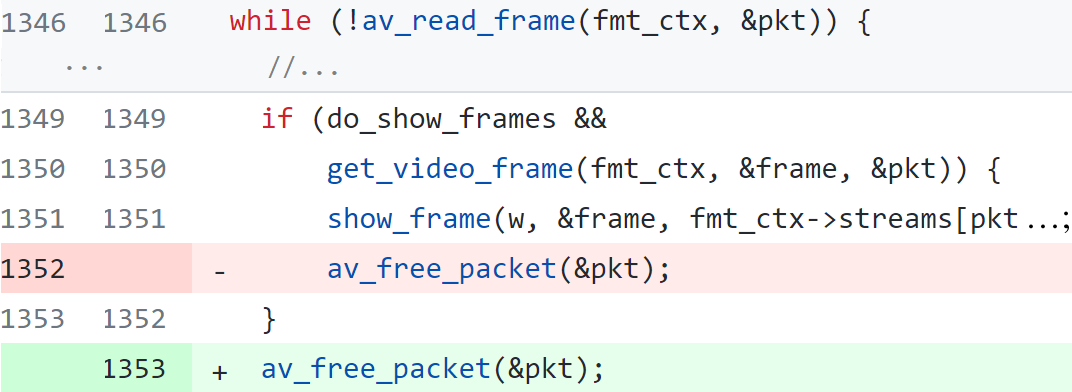}
        \caption{Commit \texttt{4fd1e2e} fixing the issue in Fig.~\ref{fig:vt2}}
        \label{fig:fixing2}
     \end{subfigure}
     
    \caption{Two commits in \textit{FFmpeg} project demonstrating the importance of considering both changed and unchanged code in understanding code changes}
\end{figure}

\textbf{Observation 1}. \textit{For a commit, the code changes' meaning is the reliable factor for deciding if the commit is dangerous or not}. 
Intuitively, a commit is dangerous to source code if it contains dangerous code changes either having vulnerabilities 
or introducing the vulnerabilities caused by the interaction between the changed code and the unchanged code.
For example, the commit in Fig.~\ref{fig:fixing-vt} contains vulnerable changed code itself, while the commit in Fig.~\ref{fig:vt2} brings a vulnerability to source code due to the relation between the changed and the unchanged code.
Specifically, in Fig.~\ref{fig:fixing-vt}, the added code itself contains a memory leak issue. 
The commit replaces the function call \texttt{avio\_read()} (at line 462, the old version) by a \texttt{if}-block examining the returned value of the call (lines 463--464, the new version). In the new version, \texttt{avio\_read()} allocates the memory of \texttt{st->codec->extradata}, yet the author forgot to free \texttt{extradata} when the values of \texttt{ret} and \texttt{len} are different (line 463 in the new version). 
In Fig.~\ref{fig:vt2}, 
adding \texttt{av\_free\_packet(\&pkt)} (line 1352, the new version) to free the packet data allocated by calling \texttt{av\_read\_frame} at line 1346 only when decoding audio frames. Consequently, the added code and the existing code collectively cause a leak problem.

Besides the code changes, to determine if a commit is suspiciously dangerous, ones could rely on expert features and commit messages~\cite{lapredict,vccfinder,jit-fine,jitline}.
Commit messages and expert features such as commit authors' experience or the number of added lines might have some relations to the suspiciousness of commits. However, they might be not necessarily discriminative for detecting dangerous commits. For example, safe commits and dangerous commits could have the same number of added lines, and be produced by the same author (e.g., the dangerous commit in Fig.~\ref{fig:fixing-vt} and the safe one in Fig.~\ref{fig:fixing}).
The application of commit messages and the expert features for JIT-VD is discussed in Sec.~\ref{sec:results}.

\begin{figure}[!h]
    \centering
    \includegraphics[width=0.75\columnwidth]{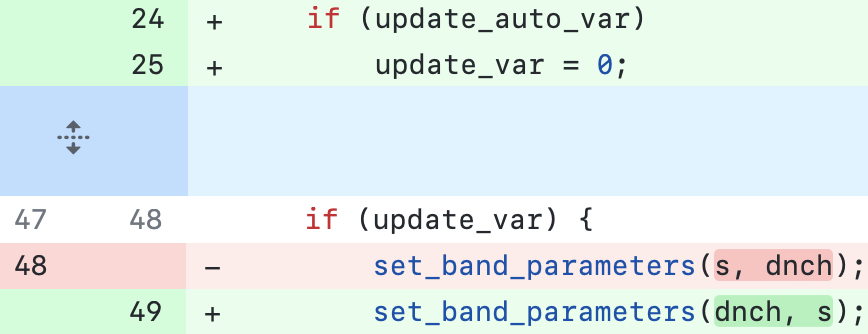}
    \caption{Unchanged code connects changed parts}
    \label{fig:ex_unchange}
\end{figure}

\textbf{Observation 2}. \textit{To precisely capture the semantics of code changes, besides the changed parts, the unchanged code is also necessary}. 
Indeed, the unchanged code could connect the related changed parts and help to understand the code changes as whole. 
In Fig.~\ref{fig:ex_unchange}, without considering the unchanged parts (e.g., line 47, the old version/line 48, the new version), the changed parts (lines 25 and 49, the new version) can be considered as semantically irrelevant. However, the added lines 24--25 affect the behavior of the added line 49 via the unchanged statement (line 48, the new one).

Additionally, once changed statements are introduced to a program, they change the program's behaviors by interacting with the unchanged statements via certain code relations such as control/data dependencies. 
%
%
Hence, precisely distinguishing similar changed parts could require their relations with the unchanged parts. 
For example, although the atomic changes adding \texttt{av\_free\_packet(\&pkt)} in Fig.~\ref{fig:vt2} and Fig.~\ref{fig:fixing2} are apparently similar, they are semantically different in their related unchanged parts. 
Adding \texttt{av\_free\_packet(\&pkt)} in Fig.~\ref{fig:vt2} is to free the packet data only when decoding audio frames, which causes a leak problem after \texttt{7328c2f}. Meanwhile, the adding one in Fig.~\ref{fig:fixing2} (line 1353, the new version) at each demuxing loop iteration, not only when decoding. This safe commit fixes the unintentional problem caused by the dangerous commit \texttt{7328c2f} in Fig.~\ref{fig:vt2}.
The suspiciousness of commit \texttt{7328c2f} (Fig.~\ref{fig:vt2}) and commit \texttt{4fd1e2e} (Fig.~\ref{fig:fixing2}) could not be determined without considering the relation between the changed code and the unchanged code. 
%

Thus, \textit{not only the changed parts but also the related unchanged parts are necessary for representing code changes to precisely capture their meaning and determine their suspiciousness}.

\textbf{Observation 3}. \textit{The changes in both the code structure and dependencies are necessary and provide rich information for understanding and determining the suspiciousness of code changes}. 
For example, in Fig.~\ref{fig:fixing2}, the memory leaking issue is fixed by the changes in the dependencies of the function called \texttt{av\_free\_packet}. Indeed, changing the call of \texttt{av\_free\_packet} from line 1352 to line 1353 (outside the \texttt{if} block) causes the call no longer control-dependent on the \texttt{if}-statement (line 1349), and creates a new control dependency of the call on the \texttt{while}-statement at line 1346. 
As a result, instead of only when decoding (inside the \texttt{if} block), \texttt{av\_free\_packet} is always called in the \texttt{while}-loop to free the packet data. Thus, it is important to capture the dependency relation of the code changes.

Additionally, the \textit{Buffer Overflow} issue shown in Fig.~\ref{fig:example_change} is brought by the changes in the code structure. Specifically, while the dependencies between the statements remain unchanged, the right-hand-side of the comparison at line 4 is changed from \texttt{BUF\_SIZE} to \texttt{2*BUF\_SIZE}. This allows line 5 to copy the amount of data larger than the capacity of \texttt{buf}. This demonstrates the importance of the structure relation in understanding the code changes.

\textit{Overall}, these observations indicate that the code changes of a commit are the \textit{deciding factor} in determining the suspiciousness of the commit. To \textit{precisely} capture the changes' meaning, not only the changed code but also the related unchanged parts are required. The changes in the code structure and dependencies are essential to \textit{comprehensively} represent the meaning of the changes.

\subsection{Key Ideas}
Based on the above observations, we introduce \tool 
centralizing the role of code changes in JIT-VD and considering the changed code of commits in relation to the unchanged code to assess commits' suspiciousness. For \tool to work, we rely on the following key ideas:

\begin{enumerate}
    
    \item We develop a graph-based representation for the code changes of commits, namely \textit{Code Transformation Graph (CTG)}. The representation captures the changes/transformations in both the code structure and dependencies of the changed code and the related unchanged code.

    \item A relational graph neural network model is employed to capture the knowledge of the known dangerous/safe commits represented in CTGs and evaluate the suspiciousness of newly-encountering commits.
    
\end{enumerate}

\section{Semantic Change Representation}

\subsection{Relational Code Graph}
From our point of view, a program is a set of \textit{inter-related elements} and \textit{relations} with other entities. 
Essentially, both the syntactic and semantic aspects of programs are necessary to properly understand the program~\cite{joern, ngo2021ranking}. 
%
In this work, we use a graph-based representation, \textit{Relational Code Graph}, for programs that represents the code elements as well as the syntactic aspect 
via the \textit{structure} relation and the semantic aspect via the \textit{dependency} relation.

\begin{figure}
    \centering
    \includegraphics[width=0.7\columnwidth]{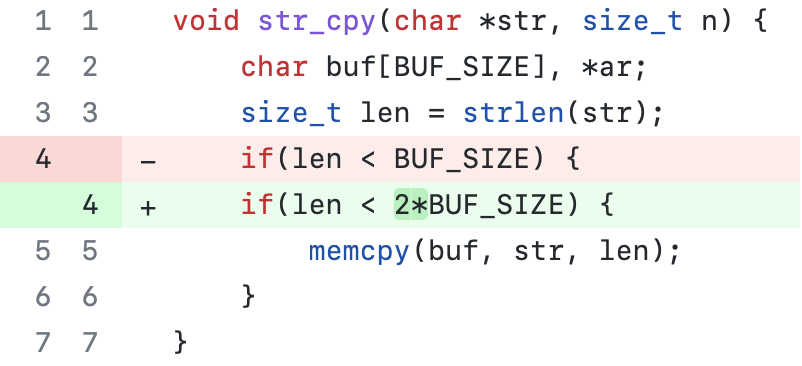}
    \caption{A dangerous commit causing a \textit{Buffer Overflow} vulnerability}
    \label{fig:example_change}
\end{figure}

\begin{figure}
    \centering
    \includegraphics[width=1.0\columnwidth]{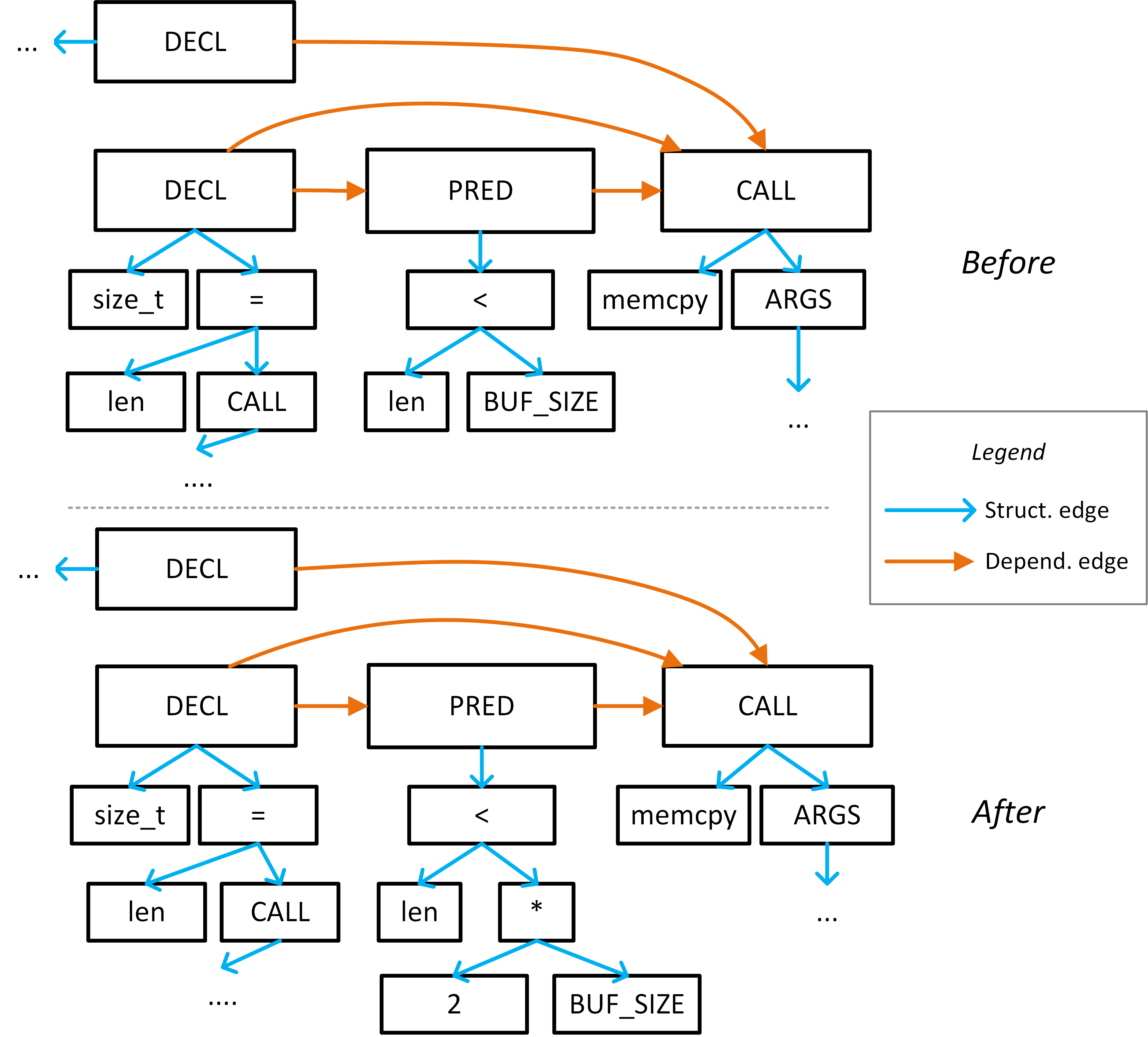}
    \caption{The relational code graphs before and after the change in Fig.~\ref{fig:example_change}}
    \label{fig:before-after}
\end{figure}

\begin{definition}[Relational Code Graph]
For a code snippet $f$, the corresponding relational code graph, a directed graph $G_f = \langle N, E, \mathcal{R} \rangle$, represents the structure and dependencies of $f$. Formally, $G_f = \langle N, E, \mathcal{R} \rangle$ is defined as followings:
\begin{itemize}
    \item $N$ is a set of nodes which are the AST nodes of $f$. In $N$, the leaf nodes are code tokens, while internal (non-leaf) nodes are abstract syntactic code elements (e.g., statements, predicate expressions, assignments, or function calls).
    
    \item $E$ is a set of edges that represent certain relations between nodes, such relations are either structure or dependency. For $n_i, n_j \in N$, an edge exists from $n_i$ to $n_j$ regarding relation $r \in \mathcal{R} = \{\textit{structure, dependency}\}$, $\exists e^r_{ij} = \langle n_i, r, n_j \rangle \in E$ if there is a relation $r$ between $n_i$ and $n_j$. 
    
\end{itemize}
\end{definition}
Note that the dependency edges connect the nodes which are statements or predicate expressions~\cite{pdg}. Each statement/predicate expression contains the descendants which are connected via the structure edges.
The reason for the incorporation of multiple kinds of knowledge about code in a single representation instead of using multiple graphs is that structural information could explain the dependencies between nodes.
For example, Fig.~\ref{fig:before-after} shows the relational code graphs of the function before and after the change in Fig.~\ref{fig:example_change}.
Statement $s$ at line 3 of Fig.~\ref{fig:example_change} assigns ($=$) a value to variable \texttt{len} which is used by statement $s'$ at line 4. There is a data-dependency relation from the statement node corresponding to $s$ to the statement node corresponding to $s'$ (Fig.~\ref{fig:before-after}).
Explicitly representing the structures of $s$ (variable \texttt{len}, assignment operator, and \texttt{strlen(str)}) and $s'$ (\texttt{len} as an argument of \texttt{memcpy}) could help explain the existence of the dependency edge from $s$ to $s'$.
In this work, we use Joern analyzer~\cite{joern} as a tool to analyze code and determine the structure and dependency relations between code elements to construct relational code graphs. 
%

\subsection{Code Transformation Graph}
\label{sec:ctg}

In this work, we aim to represent the code transformations by commits in both the code structure and the dependencies using graphs. 
As explained in Sec.~\ref{sec:example}, the changes should be represented in the context of the unchanged parts, as well as the transformations in the code elements and the relations between the elements. 

\begin{definition}[Code Transformation Graph (CTG)]
For a commit changing code from a version 
to another, 
the \textit{code transformation graph} is an annotated graph constructed from the relational code graphs of these two versions. Formally, for $G_o = \langle N_o, E_o, \mathcal{R} \rangle$ and $G_{n} = \langle N_n, E_n, \mathcal{R} \rangle$ which are the relational code graphs of the old version and the new version respectively, the CTG $\mathcal{G} = \langle \mathcal{N}, \mathcal{E}, \mathcal{R}, \alpha \rangle$ is defined as followings:
\begin{itemize}
    \item $\mathcal{N}$ consists of the code elements in the changed and unchanged parts, $\mathcal{N} = N_o \cup N_n$.

    \item $\mathcal{E}$ is the set of the edges representing the relations between nodes, $\mathcal{E} = E_o \cup E_n$.

    \item $\mathcal{R}$ is a set of the considered relations between code elements, $\mathcal{R} = \{\textit{structure, dependency}\}$.
    
    \item Annotations for nodes and edges are either \textit{unchanged}, \textit{added}, or \textit{deleted} by the change. Formally, $\alpha(g) \in \{$\textit{unchanged}, \textit{added}, \textit{deleted}$\}$, where $g$ is a node in $\mathcal{N}$ or an edge in $\mathcal{E}$:
        \begin{itemize}
            \item $\alpha(g) =$ \textit{added} if $g$ is a node and $g \in N_n \setminus N_o$, or $g$ is an edge and $g \in E_n \setminus E_o$
            \item $\alpha(g) =$ \textit{deleted} if $g$ is a node and $g \in N_o \setminus N_n$, or $g$ is an edge and $g \in E_o \setminus E_n$
            \item Otherwise, $\alpha(g) = $ \textit{unchanged} 
        \end{itemize}
\end{itemize}
\end{definition}

\begin{figure}
    \centering
    \includegraphics[width=1\columnwidth]{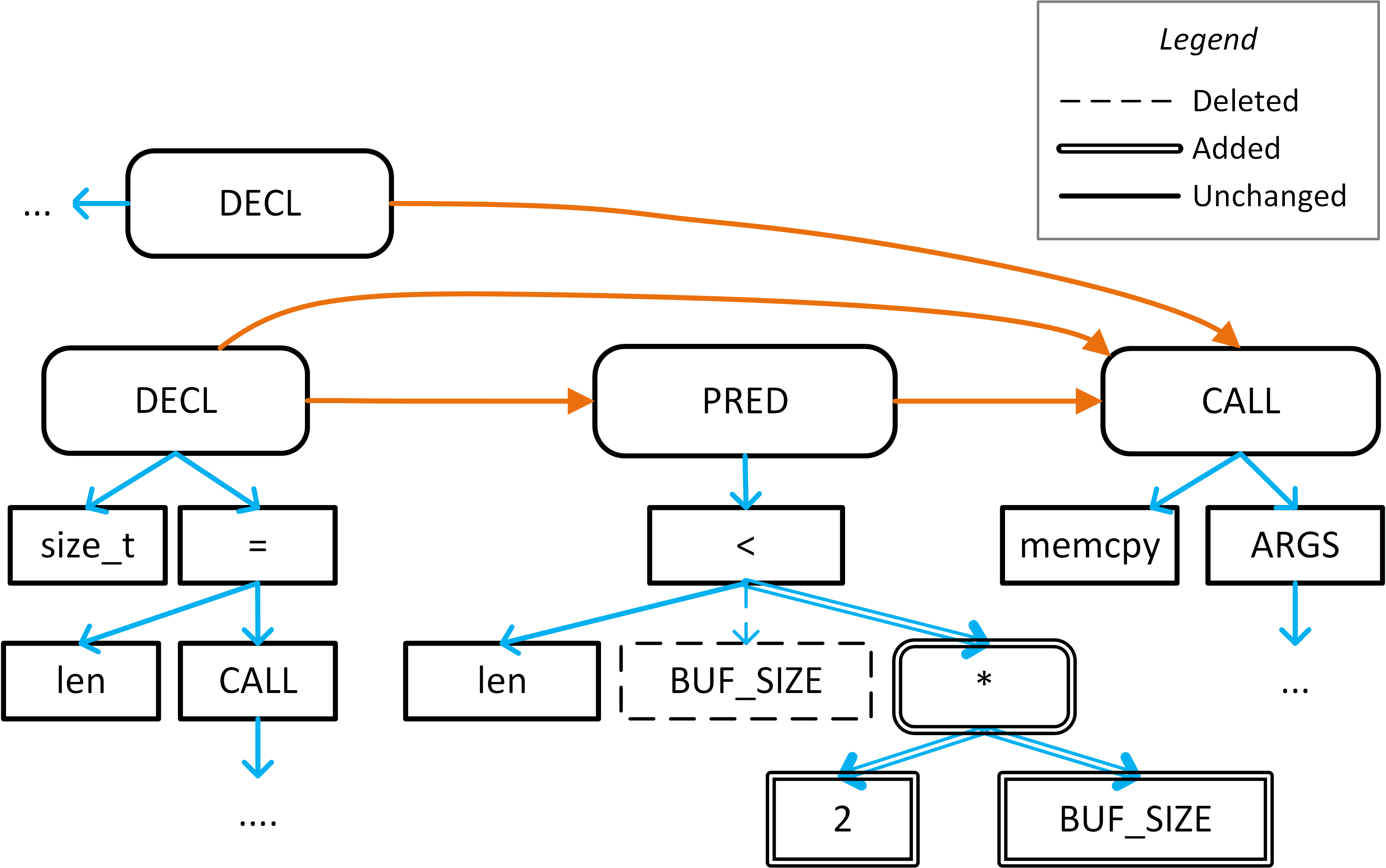}
    \caption{The CTG corresponding to the change in Fig.~\ref{fig:example_change}}
    \label{fig:example_ctg}
\end{figure}

The CTG of the change in Fig.~\ref{fig:example_change} is partially shown in Fig.~\ref{fig:example_ctg}. As we could see the structure changes, the structure of the predicate expression at line 4 is changed. Specifically, \texttt{2*BUF\_SIZE} is replaced by \texttt{BUF\_SIZE} in the right-hand side of the less-than comparison expression. This allows the statement at line 5 to be executed even when \texttt{len > BUF\_SIZE}. In the CTG, through PDG edges, the statement copies \texttt{len} (defined at line 3) bytes from \texttt{str} (declared at line 1) to \texttt{buf} (declared at line 2) which can have maximum \texttt{BUF\_SIZE} bytes or even greater. This will cause an overflow error when \texttt{2*BUF\_SIZE > len > BUF\_SIZE}.

Theoretically, a CTG corresponding to a commit could be very large and contain many unchanged nodes/edges which are irrelevant to the changes. These irrelevant parts might be unnecessary to capture the changes' meaning and also produce noise for understanding the changes. To trim those irrelevant parts, the nodes/edges that are not relevant to the changed nodes/edges are removed from $\mathcal{G}$. 
Our idea is that all the statements/predicate expressions containing changed nodes and their related statements/expressions via dependencies are kept. Additionally, the descendant nodes of these statements/expressions are also retained in the graph to enable further elaboration.
Formally, given CTG $\mathcal{G} = \langle \mathcal{N}, \mathcal{E}, \mathcal{R}, \alpha \rangle$, a node $n$ is considered as \textit{relevant} to the change if $n$ satisfies one of the following conditions:
\begin{itemize}
    
    \item $n$ is a statement/predicate expression node which is the ancestor of a deleted/added node $n'$. In other words, there is a path from $n$ to $n'$ only via structure- edges.
    
    \item $n$ is a statement/predicate expression node, and $n$ impact/is impacted by a \textit{relevant} node $n'$. In other words, there is a path from $n$ to $n'$ or from $n'$ to $n$ via dependency edges.

    \item $n$ is a descendant of a relevant node $n'$. In other words, there is a path from $n'$ to $n$ only via structure edges.
    
\end{itemize}
After trimming, the simplified CTG contains the nodes which are relevant to the changes. The edges which do not connect any pair of nodes are also removed from the graph.

\section{Code-centric Just-In-Time Vulnerability Detection with Graph Neural Networks}

\begin{figure*}
    \centering
    \includegraphics[width=2.0\columnwidth]{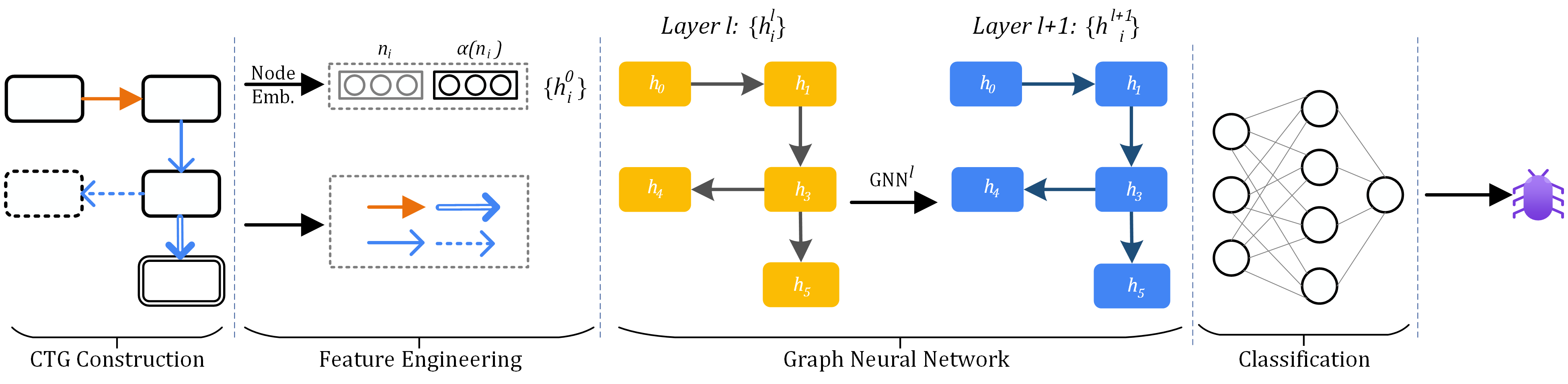}
    \caption{Just-In-Time Vulnerability Detection Model in \tool}
    \label{fig:model}
\end{figure*}

Fig.~\ref{fig:model} shows the overview of our JIT-VD model. In this work, each CTG is treated as a relational graph. 
%
%
%
To capture the important features of the graphs, not only the contents of the nodes but also their relations need to be considered in JIT-VD. 
For this purpose, we apply relational graph neural networks~\cite{rgat, rgcn} for the JIT-VD task.
%

First, the nodes are embedded into $d$-dimensional hidden features $n_i$ produced by embedding the content of the nodes. 
Particularly, to build the vectors for nodes' content, we use Word2vec~\cite{word2vec_1} which is widely used to capture semantic similarity among code tokens~\cite{embedding_emse22}.
Then, to form the node feature vectors, the node embedding vectors are annotated with the change operators (\textit{added}, \textit{deleted}, and \textit{unchanged}) by concatenating corresponding one-hot vector of the operators to the embedded vectors, $h^{0}_i = n_i \bigoplus \alpha(n_i)$, where $\bigoplus$ is the concatenating operator and $\alpha$ returns the one-hot vector corresponding the annotation of node $i$.
The resulting vectors are fed to a relational graph neural network model which could be a relational graph convolution network (RGCN)~\cite{rgcn} or a relational graph attention (RGAT) model~\cite{rgat}.
Each layer of RGCN or RGAT computes the representations for the nodes of the graph through message passing, where each node gathers features from its neighbors under every relation to represent the local graph structure. Stacking $L$ layers allows the network to build node representations from the $L$-hop neighborhood of each node.
%
%
Thus, the feature vector $h^{l+1}_{i}$ at node $i$ at the next layer is: 
\begin{equation} \label{eq:propagate_rule}
h^{l+1}_{i} =\sigma \left( 
\sum_{r \in \mathcal{R}}
    \sum_{j \in \mathcal{N}^r_i} \frac{1}{c_{i,r}} g_j^r
\right)
\end{equation}
%
%
%
%
%
%
%
%
where $g_j = W^l_r h^{l}_{j}$ is the distinct intermediate representation of node $j$ under relation $r$, $W^l_r$ is a learnable weight matrix for feature transformation specific for relation $r$. In equation (\ref{eq:propagate_rule}), $\mathcal{N}^r_i$ is the set of neighbor indices of node $i$ under relation $r \in \mathcal{R} = \{\textit{structure, dependency}\}$.
%
Additionally, $1/c_{i}^r$ is a problem-specific normalization constant that can be chosen in advance (such as $c_{i}^r = |\mathcal{N}_i^r|$). 
%
Meanwhile, $\sigma$ is a non-linear activation function such as ReLU.
In this work, instead of using constant $1/c_{i}^r = |\mathcal{N}_i^r|^{-1}$, we replace the normalization constant by an learnable attention weight $a_{ij}^r$, where $\sum_{j,r} a_{ij}^r = 1$, to allow the model learn the importance of node $j$ and the relation between node $i$ and node $j$ under relation $r$~\cite{rgat}:
%
%
$$
a_{ij}^r = \text{softmax}_{j}(E_{ij}^r) =  \frac{\exp{(E_{ij}^r)}}{\sum_{k \in \mathcal{N}^r_i} \exp{(E_{ik}^r)}} 
$$
$$
E_{ij}^r = \text{LeakyReLu}(q_i^r + k_j^r)
$$
$$
q_i^r = g_i^r Q^r, \text{ and } k_j^r = g_j^r K^r
$$
where $Q^r$ and $K^r$ are the query kernel and key kernel projecting $g_i$ (and $g_j$) into query and key representations~\cite{rgat}. $Q^r$ and $K^r$ are combined to form the attention kernel for relation $r$, $A^r = Q^r \bigoplus K^r$.

%
After $L$ GNN layers, a $d$-dimensional graph-level vector representation $H$ for the whole CTG $\mathcal{G} = \langle \mathcal{N}, \mathcal{E}, \alpha \rangle$ is built by aggregating over all node features in the final GNN layer, $H = \Phi_{i \in [1, |\mathcal{N}|]} h^L_i$, where $\Phi$ is a graph readout function such as \textit{sum}, \textit{average}, or \textit{max}. The impact of the aggregation function $\Phi$ on \tool's performance will be empirically shown in Sec.~\ref{sec:results}.
Finally, the graph features are then passed to a Multilayer perceptron (MLP) to classify if $\mathcal{G}$ is dangerous or not.

\section{Evaluation Methodology}
\label{sec:eval}
To evaluate our \jit vulnerability detection approach, we seek to answer the following research questions:

\noindent\textbf{RQ1: \textit{Accuracy and Comparison}.} How accurate is {\tool} in detecting dangerous commits? And how is it compared to the state-of-the-art approaches~\cite{vccfinder,cc2vec,codereviewer}?

\noindent\textbf{RQ2: \textit{Intrinsic Analysis}.} How do the JIT-VD model's properties/components, including the GNN model, the number of GNN layers, and the graph readout function in {\tool} impact \tool's performance?

\noindent\textbf{RQ3: \textit{Change Representation Analysis}.} How do related unchanged parts in CTGs impact {\tool}'s performance? And, how do the structure relation and dependency relation in CTGs impact {\tool}'s performance?

\noindent\textbf{RQ4: \textit{Sensitivity Analysis}.} How do various the input's factors, including training data size and changed code's complexity affect {\tool}'s performance?

\noindent\textbf{RQ5: \textit{Time Complexity}.} What is  {\tool}'s running time?

\subsection{Dataset}
\label{sec:data}
To facilitate applying advanced ML techniques to automatically and effectively learn latent and abstract dangerous commit patterns, we collected a large number of dangerous and safe commits in numerous real-world projects based on the well-known SZZ algorithm~\cite{sliwerski2005changes}. 
We also improved the process of collecting dangerous and safe commits in the existing work~\cite{vccfinder,vuldigger,are_we_there}.
%

%
%
Particularly, to identify \textit{dangerous commits}/\textit{vulnerability contributing commits (VCCs)}~\cite{vccfinder}, the general idea is to start from vulnerable statements and identify the previous commits which last modify these statements\footnote{This can be done by using \texttt{git blame}}. 
Specifically, the vulnerable statements are identified from the \textit{vulnerability-fixing} commits based on the following heuristics. 
The statements which are deleted by the fixing commits could be considered to be vulnerable~\cite{vccfinder,vuldigger, vul_secret, bigvul}. 
%
%
%
For the statements added by the fixing commits, the existing studies~\cite{vccfinder,vuldigger} assume that such statements are added to fix the nearby statements. In particular, the statements which are \textit{physically} surrounding the added statements are considered to be vulnerable regardless of whether these statements semantically relate to added statements or not. 
As a result, this procedure could miss the actual VCCs and instead blame commits which are irrelevant. To reduce this risk, we blame the statements which are \textit{semantically} related to the added statements via data/control dependencies rather than the physically surrounding ones. 


%
%
In practice, forming a vulnerability in a code version could require one or more VCCs over time. 
However, the existing studies~\cite{jit_vul_survey, fan2019impact} illustrate that various of VCCs are false positives due to refactoring operators or tangling problems of the fixing commits. For collecting VCCs, although multiple improvements of SZZ have been proposed, this is still an inevitable issue~\cite{fan2019impact}. Moreover, warning developers by providing any potential VCCs could frustrate them and make the developers spend undesirable efforts. 
Thus, in this work, to reduce noises while still maintaining the purpose of detecting vulnerabilities as soon as before it is merged to source code, we consider \textit{vulnerability triggering commits} as \textit{dangerous} ones which are the last VCCs triggering to fully expose vulnerabilities. 
For a commit irrelevant to a known vulnerability, it might still contain unknown vulnerabilities~\cite{vccfinder,are_we_there,vuldigger}.
Meanwhile, the fixing commits had already fixed some problems, been reviewed, and reported to the community. Thus, it could be reasonable to consider the fixing commits, which are not a VCC, as \textit{safe commits}.

In this work, we extract the vulnerability-fixing commits from various public vulnerability datasets~\cite{fixing_database1,bigvul,devign}.
In total, we collected 20,274 commits including 11,299 safe commits and 8,975 dangerous commits from the vulnerabilities reported from Aug 1998-Aug 2022 in real-world 506 C/C++ projects such as FFmpeg, Qemu, Linux, and Tensorflow. Table~\ref{tab:dataset} shows the overview of our dataset. The details of our dataset can be found at: \textit{\url{https://github.com/ttrangnguyen/CodeJIT}}.

\begin{table}[]
\centering
\caption{Dataset statistics}
\label{tab:dataset}
\begin{tabular}{@{}lrrrr@{}}
\toprule
           & \begin{tabular}[c]{@{}r@{}}\#\textbf{Safe} \\ \textbf{commits}\end{tabular} & \begin{tabular}[c]{@{}r@{}}\#\textbf{Dangerous} \\ \textbf{commits}\end{tabular} & \textbf{\%adds} & \begin{tabular}[c]{@{}l@{}}$|N|/|E|$ \\ of \textbf{CTGs}\end{tabular} \\
           \midrule
\textit{Ffmpeg}     & 4,449 &      3,462       &     73.95               & 0.70                       \\
\textit{Qemu}     &  3,551 &      3,183       &       76.80            &  0.68                    \\
\textit{Linux}    &   783   &    780      &         78.45           &      0.66                 \\
\textit{Tensorflow} &  224   &    189       &          81.48          &  1.18                   \\
\multicolumn{5}{c}{\textit{502 projects more...}}                        \\ \midrule
\textit{Total}      &  11,299     &    8,975   &         73.31           &   0.70                     \\ \bottomrule
\end{tabular}
\end{table}

\begin{table}[]
\centering
\caption{Two evaluation settings: \dev and \cross}
\label{tab:setting-details}
\begin{tabular}{@{}clrrr@{}}
\toprule
\textbf{Setting}                        &          & \begin{tabular}[c]{@{}r@{}}\#\textbf{Dangerous} \\ \textbf{commits}\end{tabular} & \begin{tabular}[c]{@{}r@{}}\#\textbf{Safe} \\ \textbf{commits}\end{tabular} & \#\textbf{Commits} \\ \midrule
\multirow{2}{*}{\textit{Dev.-process}}  & Training & 7,748 & 8,471 & 16,219    \\
                                        & Testing  & 1,227 & 2,828 & 4,055     \\ \midrule
\multirow{2}{*}{\textit{Cross-project}} & Training & 7,714 & 9,176 & 16,890    \\
                                        & Testing  & 1,261 & 2,123 & 3,384     \\ \bottomrule
\end{tabular}
\end{table}

\subsection{Evaluation Setup, Procedure, and Metrics}

\subsubsection{Empirical Procedure} 

\textbf{RQ1. Accuracy and Comparison}. 

\textbf{\textit{Baselines.}} We compared \tool against the state-of-the-art JIT-VD approaches and the advanced approaches for just-in-time general defect detection adapted for JIT-VD: 

1) \textbf{VCCFinder}~\cite{vccfinder}, which uses 
commit messages and expert features to train a Support Vector Machine and is specialized for the JIT-VD task. 

2) \textbf{CC2Vec}~\cite{cc2vec} + \textbf{DeepJIT}~\cite{deepjit}: A CNN-based just-in-time defect prediction that the features of commits are captured by CC2Vec, a pre-trained model vectorizing commits using their changed code and commit message; 

3) \textbf{JITLine}~\cite{jitline}: A simple but effective method utilizing changed code and expert features to detect buggy commits; 

4) \textbf{LA\_Predict}~\cite{lapredict}: A regression-based approach simply using the
added-line-number feature which can outperform CC2Vec and DeepJIT; and

5) \textbf{JITFine}~\cite{jit-fine}: A DL-based approach extracting features of commits from changed code and commit message using CodeBERT as well as expert features.

\textit{\textbf{Procedure.}} In this comparative study, we evaluate the performance of the approaches in two real-world settings used in the existing studies~\cite{lapredict,jit-fine,jitline,vccfinder,cross-setting,deepjit}: \textit{development-process} and \textit{cross-project}. The details of data splitting for both settings are shown in Table~\ref{tab:setting-details}.

In the \textit{development- (dev.-) process} setting considering the impact of time on the JIT-VD approaches' performance, we follow the same time-aware evaluation procedure to construct the training data and testing data from the dataset as in the prior work~\cite{jit-fine,jitline,vccfinder,deepjit}.
Particularly, we divided the commits into those before and after time point $t$. The dangerous/safe commits before $t$ were used for training, while the commits after $t$ were used for evaluation. We selected a time point $t$ to achieve a random training/test split ratio of 80/20 based on time. Specifically, for the \textit{dev.-process} setting, the commits from Aug 1998 to Mar 2017 are used for training and the commits from Apr 2017 to Aug 2022 are for evaluation. In total, the training/test split in the number of commits for this setting is 16,219/4,055.

Similar to existing work~\cite{cross-setting,lapredict}, in the \textit{cross-project} setting, we evaluate how well the approaches can learn to recognize dangerous commits in a set of projects and detect suspicious commits in the other set. 
%
Specifically, the commits in a fixed set of projects are used to train the approaches, the remaining set of commits is used for testing. 
For this setting, the whole set of projects is randomly split into 80\% (402 projects) for training and 20\% (104 projects) for testing. The training/test split in the number of commits for this setting is 16,890/3,384.

\textbf{RQ2. Intrinsic Analysis}. We also investigated the impact of the GNN model and the number of GNN layers on \tool's JIT-VD performance. We used different variants of relational graph neural networks, and their properties to study the impact of those factors on \tool's performance.

\textbf{RQ3. Change Representation Analysis}. We studied the impacts of the code relations (structure and dependency) and unchanged code in CTGs on \tool's performance. We used different variants of CTGs and measured the performance of \tool with each variant.

\textbf{RQ4. Sensitivity Analysis}. We studied the impacts of the following factors on the performance of \tool: training size and change complexity (the rate of the changed nodes over the total of nodes in CTGs). To systematically vary these factors, we gradually added more training data, and varied the range of the change rate.

\subsubsection{Metrics}
To evaluate the JIT-VD approaches, we measure the classification \textit{accuracy}, \textit{precision}, and \textit{recall}, as well as \textit{F1} which is a harmonic mean of precision and recall. 
Particularly, the classification accuracy (\textit{accuracy} for short) is the fraction of the (dangerous and safe) commits which are correctly classified among all the tested commits.
For detecting dangerous commits, \textit{precision} is the fraction of correctly detected dangerous commits among the detected dangerous commits, while \textit{recall} is the fraction of correctly detected dangerous commits among the dangerous commits. Formally $precision = \frac{TP}{TP+FP}$ and $recall = \frac{TP}{TP+FN}$, where $TP$ is the number of true positives, $FP$ and $FN$ are the numbers of false positives and false negatives, respectively. \textit{F1} is calculated as $\textit{F1} = \frac{2 \times precision \times recall}{precision + recall}$.

\section{Experimental Results}
\label{sec:results}

\subsection{JIT-VD Performance Comparison (RQ1)}
Table~\ref{tab:vd-perf-time} and Table~\ref{tab:vd-perf-proj} show the performance of \tool and the other JIT-VD approaches in the \dev and \cross settings. \tool significantly outperforms the state-of-the-art JIT-VD approaches in JIT-VD \textit{F1} and classification \textit{accuracy} in both settings.

\begin{table}[]
\centering
\caption{\textit{Dev.-process:} JIT-VD performance comparison}
\label{tab:vd-perf-time}
\begin{tabular}{@{}l|rrr|r@{}}
\toprule
                   & \textit{Precision} & \textit{Recall} & \textit{F1}   & \textit{Accuracy} 
                   \\ \midrule
VCCFinder          & 0.50	            & \textbf{0.75}	        & 0.60	          & 0.70           \\
CC2Vec+DeepJIT     & 0.33               & 0.66          & 0.44            & 0.49            \\
LA\_Predict        & 0.59               & 0.58          & 0.59            & 0.75            \\
JITLine            & 0.62	            & 0.66	        & 0.64	          & 0.77	        \\
JITFine            & 0.62	            & 0.73	        & 0.67	          & 0.78	        \\
\tool              & \textbf{0.78}	    & 0.70	        & \textbf{0.74}	  & \textbf{0.85}	\\ \bottomrule
\end{tabular}
\end{table}

\begin{table}[]
\centering
\caption{\textit{Cross-project}: JIT-VD performance comparison}
\label{tab:vd-perf-proj}
\begin{tabular}{@{}l|rrr|r@{}}
\toprule
                   & \textit{Precision}  & \textit{Recall}    & \textit{F1}    & \textit{Accuracy}  \\ \midrule
VCCFinder          & 0.58	    & 0.49	    & 0.53	& 0.74          \\
CC2Vec+DeepJIT     & 0.40	    & 0.69	    & 0.51	& 0.50	    \\
LA\_Predict        & 0.66	    & 0.48	    & 0.56	& 0.72	    \\
JITLine            & 0.57	    & 0.72	    & 0.64	& 0.69	    \\
JITFine            & 0.68	    & 0.69	    & 0.68	& 0.76	    \\
\tool              & \textbf{0.90}	& \textbf{0.80}	& \textbf{0.84}	& \textbf{0.89} \\ \bottomrule
\end{tabular}
\end{table}

In the \dev setting, the JIT-VD \textit{F1} and \textit{accuracy} of \tool are significantly better than those of the baseline approaches by \textbf{10--68\%} and \textbf{8--73\%}, respectively. Meanwhile, \tool's \textit{recall} in the \dev setting is just slightly lower than those of VCCFinder~\cite{vccfinder} and JITFine~\cite{jit-fine}, by up to 7\%. The predictions of \tool are much more precise than those of all the studied methods. Specifically, \tool improves the \textit{precision} of the other approaches by \textbf{15--136\%} in this setting.

In the \cross setting, \tool consistently improves the state-of-the-art JIT-VD approaches by \textbf{23--65}\% in JIT-VD \textit{F1} and \textbf{17--77}\% in classification \textit{accuracy}.
\tool achieves the \textit{recall} of \textbf{0.8}, which is significantly better than the corresponding rates achieved by all the other approaches, with improvements ranging from \textbf{11--66\%}. This means that \tool can effectively detect \textbf{8/10} dangerous commits.
Especially, the \textit{precision} of \tool in this setting is \textbf{90\%}. In other words, \textbf{9/10} suspicious commits detected by \tool actually are dangerous ones. Meanwhile, the corresponding figures of the existing methods are 4.0--6.8.

Moreover, the results in Table~\ref{tab:vd-perf-time} and Table~\ref{tab:vd-perf-proj} show that \tool performs stably in both settings. Particularly, \tool consistently achieves the highest overall performance (\textit{F1} and \textit{accuracy}) among the JIT-VD approaches. \tool also maintains the same pattern of improvement for both \textit{precision} and \textit{recall} when switching between the \dev setting and \cross setting. 
For most of the other approaches, an increase in \textit{precision} is typically accompanied by a decrease in \textit{recall} when switching from one setting to the other. For example, in the \dev setting, VCCFinder can find much more dangerous commits (higher \textit{recall}), yet raise more false alarms (lower \textit{precision}) compared to this approach in the \cross setting. In other words, these approaches cannot detect dangerous commit more effectively and more precisely at the same time.


\begin{gtheorem}
\textbf{Answer for RQ1}: For both settings, \tool can precisely detect a larger number of dangerous commits compared to the state-of-the-art approaches. This confirms our code-centric strategy in detecting vulnerabilities at the commit-level. 
\end{gtheorem}

\begin{figure}
     \centering
     \begin{subfigure}[b]{0.5\textwidth}
        \centering
        \includegraphics[width=0.7\columnwidth]{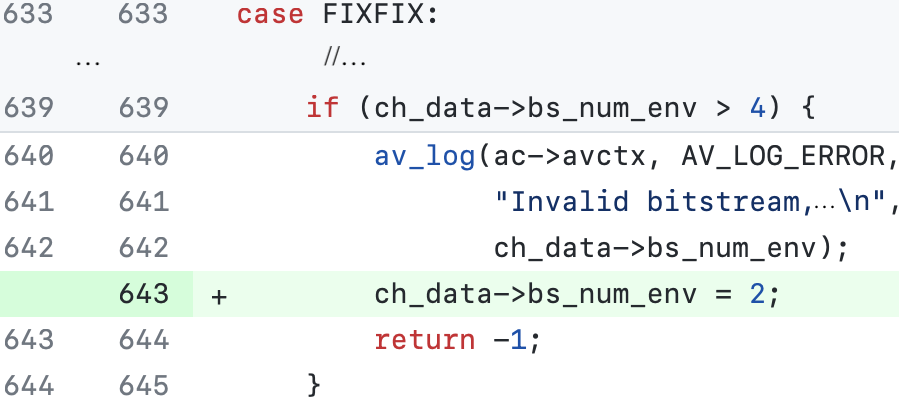}
        \caption{Commit \texttt{a8ad83b} fixing a vulnerability in \textit{FFmpeg} but causing another} 
        \label{fig:RQ1_EX1}
     \end{subfigure}
     \hfill

     \begin{subfigure}[b]{0.5\textwidth}
        \centering
        \includegraphics[width=0.85\columnwidth]{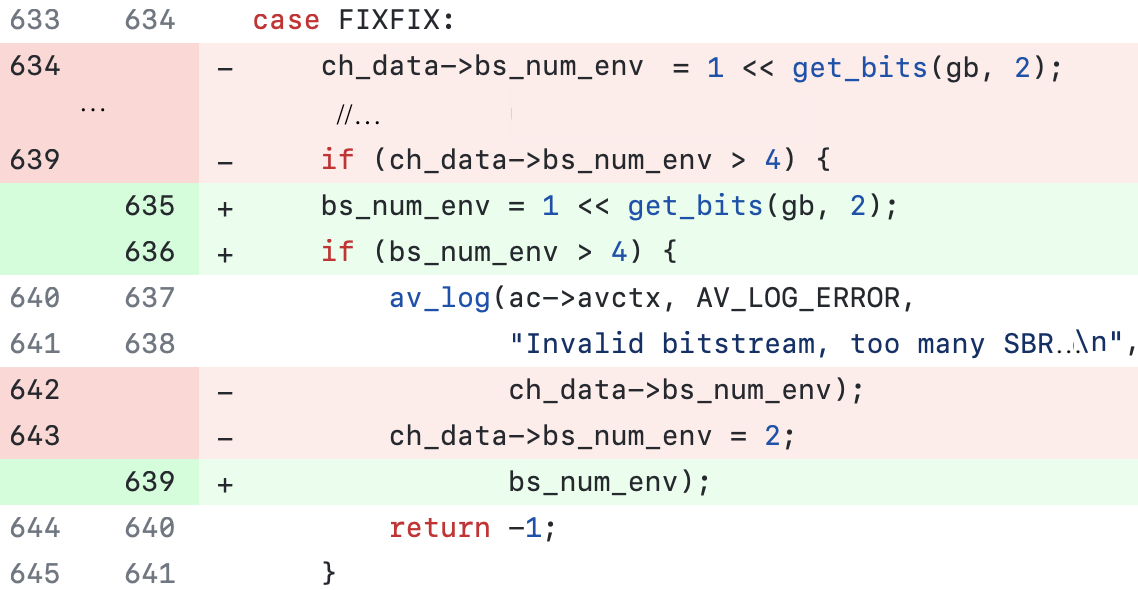}
        \caption{Commit \texttt{87b08ee} fixing the vulnerability in Fig.~\ref{fig:RQ1_EX1}}
        \label{fig:RQ1_EX2}
     \end{subfigure}
     
    \caption{Example of a dangerous commit correctly detected by \tool}
\end{figure}

\textbf{Result Analysis.} 
%
%
%
%
%
%
%
%
In our experiments, we found that 69/188 fixing commits, which cause other issues, are correctly detected by \tool. Meanwhile, the corresponding figures of JITLine and LA\_Predict are 58 and 41, respectively. 
Fig.~\ref{fig:RQ1_EX1} shows a commit fixing a bug, yet causing another problem in \textit{FFmpeg}. Specifically, to prevent the out-of-bound problem, \texttt{ch\_data->bs\_num\_env} is set to a valid value if it is larger than the allowance (line 643 in the new version, Fig.~\ref{fig:RQ1_EX1}). However, changing the value of this variable without appropriately checking causes another serious problem, which is later fixed by the commit \texttt{87b08ee}. 
%
The dangerous commit in Fig.~\ref{fig:RQ1_EX1} is correctly detected by \tool, while the other methods misclassify this commit as a \textit{safe} one.

Specifically, the approaches leveraging commit message such as JITLine and JITFine could be misled by special keywords such as ``\textit{Fixes}'' in the commit message:
``\textit{avcodec/aacsbr\_template: Do not leave bs\_num\_env invalid \textbf{Fixes} out of array read \textbf{Fixes}: 1349/clusterfuzz...}''.
Zeng~\etal~\cite{lapredict} show that the message of a commit could provide the intention of the commit.
However, developers often unintentionally bundle unrelated changes with different purposes (e.g., bug fix and refactoring) in a single commit (tangled commit)~\cite{untangling}.  
Multiple studies have shown that fixing a bug could cause another bug~\cite{guo2010characterizing, purushothaman2005toward}. 
%
Moreover, in practice, commit messages are often poor-quality, even empty~\cite{good_commit_message}. This makes the application of commit messages for JIT-VD much less reliable and causes incorrect predictions.
Meanwhile, LA\_Predict considers only the line-added-number to evaluate the suspiciousness based on the hypothesis that the larger the line-added-number, the higher the probability of being defective~\cite{lapredict}. In fact, the commit in the example above contains only two added lines of code, which leads to the incorrect prediction of LA\_Predict. 

By capturing the changes in the code structure expressed in the commit's CTG, \tool is able to recognize that the value of \texttt{ch\_data->bs\_num\_env} is dangerously changed (line 643 of the new version) after checking the allowed bound (line 639). In addition, by analyzing the dependencies, \tool can identify that the newly changed value is a magic number (\texttt{2}) and has no relation with the checked bound (\texttt{4}), which means that the corresponding commit still does not guarantee the safety of code after being executed. 
Then, this issue is fixed by commit \texttt{87b08ee} (Fig.~\ref{fig:RQ1_EX2}) by using a temporary variable (\texttt{bs\_num\_env}) instead of directly using and defining \texttt{ch\_data->bs\_num\_env}. 
As expected, \tool  correctly detects both these two commits.
%
%



Overall, \textit{properly utilizing the code-related aspects of commits could help} \tool \textit{precisely understand the semantics of the changes and effectively learn to differentiate safe/dangerous commits, thus significantly improves the JIT-VD performance}.

\subsection{Intrinsic Analysis (RQ2)}
\label{sec:intrinsic}

\subsubsection{Impact of GNN Model}
To investigate the impact of different GNN models on the JIT-VD performance, we compare three variants of relational graph networks: \tool with RGCN~\cite{rgcn},  RGAT~\cite{rgat},  and  FastRGCN~\cite{fastgcn}. 
In this experiment, we used the \textit{sum} readout function and two GNN layers, and applied for the \dev setting.
The results of those three variants is shown in Table~\ref{tab:gnn_models}. As expected, \tool obtains quite stable performance with about 0.74\% in \textit{F1} and 0.85 in \textit{accuracy}.
Moreover, while \tool obtains quite similar \textit{precision} and \textit{recall} with both RCGN and FastRGCN, it archives the highest \textit{precision}, yet lowest \textit{recall} with RGAT.
Indeed, the attention mechanism enables RGAT to focus on the important features of the neighboring nodes in each relation for computing the features of the graph nodes. This helps RGAT precisely detect dangerous commits and obtains the highest \textit{precision}. However, 
the attention mechanism in the RGAT model may be causing it to prioritize certain features, leading to higher \textit{precision} but lower \textit{recall}. This can happen if the attention mechanism is too focused on specific features and fails to capture the broader context of the data.

\begin{table}[]
\centering
\caption{Impact of GNN models on JIT-VD Performance}
\label{tab:gnn_models}
\begin{tabular}{@{}l|rrr|r@{}}
\toprule
                    & \textit{Precision} & \textit{Recall}   & \textit{F1}   & \textit{Accuracy}  \\ \midrule
RGCN  & 0.78	            & 0.70	            & 0.74	        & 0.85           \\
FastRGCN     & 0.77               & 0.71          & 0.74            & 0.85       \\
RGAT     & 0.82               & 0.66          & 0.73            & 0.85           \\
\bottomrule
\end{tabular}
\end{table}

\subsubsection{Impact of Number of GNN Layers}

\begin{figure}[!h]
    \centering
    \includegraphics[width=0.9\columnwidth]{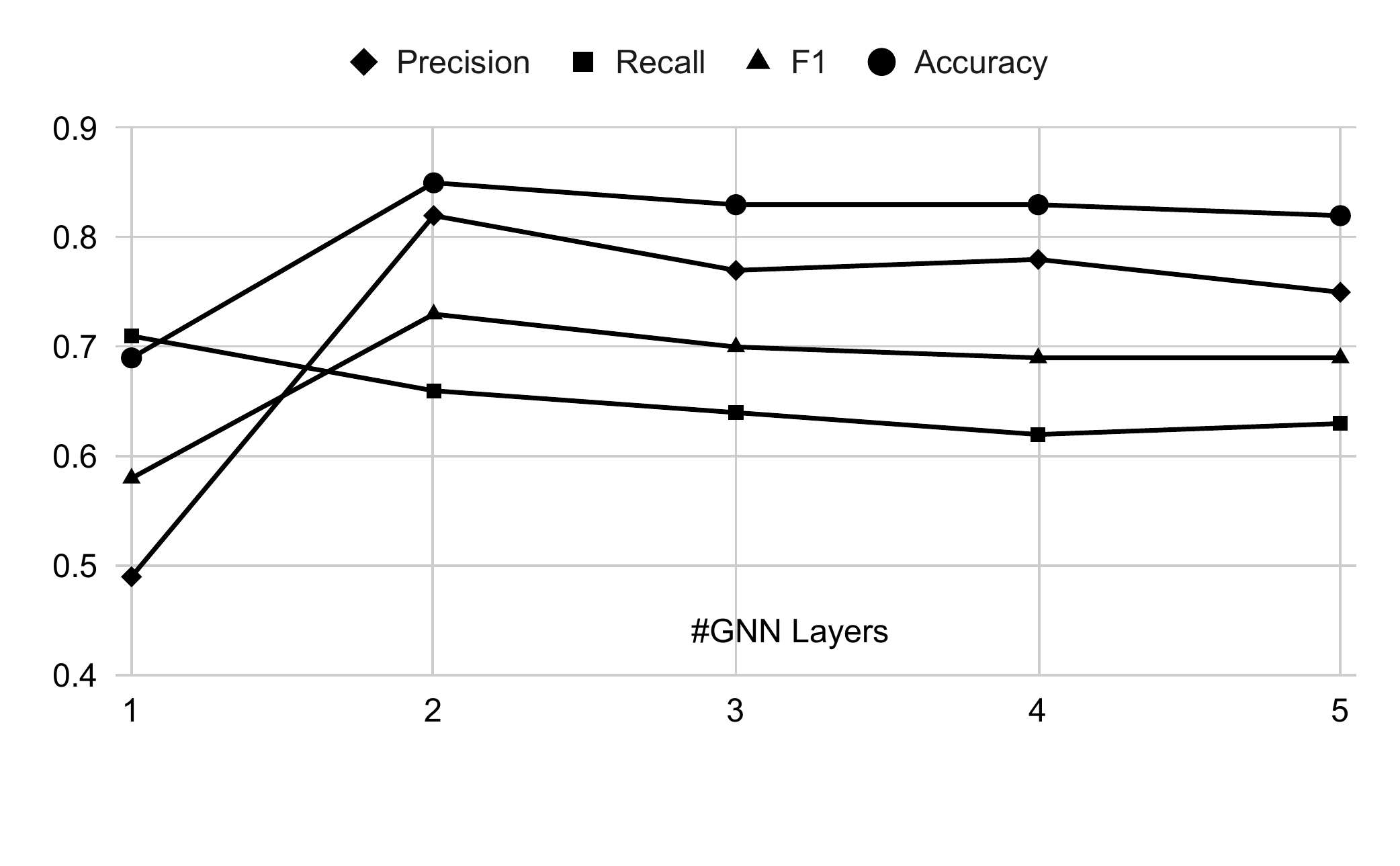}
    \caption{Impact of the number of GNN layers}
    \label{fig:impact_num_layers}
\end{figure}

To study the impact of the number of GNN layers on the JIT-VD performance of \tool, we varied the number of GNN layers in JIT-VD model from 1 to 5. 
In this experiment, we used RGAT for the JIT-VD model and the \dev setting.
As seen in Fig.~\ref{fig:impact_num_layers}, the JIT-VD \textit{F1} and classification \textit{accuracy} of \tool are significantly better when increasing the number of RGCN layers in the JIT-VD model from one to two. With two RGCN layers, the model represents each node in CTGs from the 2-hop neighborhood. Meanwhile, each node is represented by considering only its neighbors by the model with one RGCN layer.
Thus, by using more information to represent each node in CTGs, the model detected dangerous commits more precisely. Indeed, the precision of \tool is  remarkably improved by 67\% with two RGCN layers. As expected, the more complex node representation mechanism slightly lowers the recall by 7\%.
However, considering longer 2-hop neighborhood to represent nodes in CTGs causes downgrades in the detection performance. Particularly, the performance \tool gracefully decreases when the number of GNN layers increases from 2 to 5. This could be because for a node, the further information collected from the other nodes could bring noises in representing the node. Additionally, with a larger number of layers and a longer neighborhood relation, the set of the shared neighbors which are used to represent nodes could be larger. Thus, the different nodes could be represented similarly. Consequently, the discriminating performance of GNN model is reduced.
Moreover, the model with more layers is much more complicated. Thus, we use two GNN layers to ensure the best performance and simplicity.

\subsubsection{Impact of Graph Readout Function}

To evaluate the impact of the different readout functions in forming the whole feature vector of a graph, we create different variants of \tool with \textit{Max}, \textit{Average}, and \textit{Sum} function to aggregate node features. 
In this experiment, we used the RGAT model, with two layers for the \cross setting.
As shown in Table~\ref{tab:impact_readout_function}, \tool's performance is not significantly affected when the aggregation function is changed. Specifically, the average \textit{F1} and \textit{accuracy} are 0.81 and 0.88, respectively.

\begin{table}[]
\centering
\caption{Impact of graph readout function}
\label{tab:impact_readout_function}
\begin{tabular}{@{}l|rrr|r@{}}
\toprule
                    & \textit{Precision} & \textit{Recall}   & \textit{F1}   & \textit{Accuracy}  \\ \midrule
\textit{Max}  & 0.93	            & 0.73	            & 0.81	        & 0.88           \\
\textit{Average}     & 0.92               & 0.73          & 0.81            & 0.88           \\
\textit{Sum}     & 0.95               & 0.70          & 0.80          & 0.87       \\

\bottomrule
\end{tabular}
\end{table}

\begin{gtheorem}
\textbf{Answer for RQ2}: The JIT-VD model's properties/components impact \tool's effectiveness differently. The GNN model and graph readout function slightly impact the performance, while the number of GNN layers could significantly impact \tool's performance.
\end{gtheorem}
\subsection{Change Representation Analysis (RQ3)}
\subsubsection{Unchanged Code's Role Analysis}
To investigate the contribution of the related unchanged code in CTGs, we used two variants of CTG: one considering both changed code and unchanged code (CTG), the other considering only changed code ($\hat{\text{CTG}}$). Table~\ref{tab:unchanged-code} shows the JIT-VD performance of \tool using the two CTG variants: $\tool_\text{CTG}$ and $\tool_{\hat{\text{CTG}}}$. 
Note that, in this experiment, we applied for the \cross setting and used the same RGAT model, with two layers, and sum readout function for both CTG variants.

As seen, additionally considering the related unchanged code along with changed code in CTGs significantly improves the JIT-VD performance of \tool using CTGs with only changed code. Particularly, \tool achieves better \textit{precision} and \textit{recall} by 20\% and 15\%, respectively. 
The related unchanged code provides valuable information and helps the model not only understand code changes more precisely but also discover more vulnerability patterns. 
The number of dangerous commits detected by $\tool_\text{CTG}$ but not detected by $\tool_{\hat{\text{CTG}}}$ (149 commits) nearly triples the corresponding figure by $\tool_{\hat{\text{CTG}}}$ but not detected by $\tool_\text{CTG}$ (only 52 commits).
This experimentally confirms our observation 2 on the important role of related unchanged code for understanding code changes in JIT-VD.

\begin{table}[]
\centering
\caption{Impact of the unchanged code in CTGs on \tool's performance}
\label{tab:unchanged-code}
\begin{tabular}{l|rrr|r}
\toprule
                       & \textit{Precision} & \textit{Recall} & \textit{F1} & \textit{Accuracy} \\ \midrule
$\tool_\text{CTG}$    & 0.95               & 0.70            & 0.80        & 0.87              \\
$\tool_{\hat{\text{CTG}}}$ & 0.79               & 0.61            & 0.69        & 0.79              \\ \bottomrule
\end{tabular}
\end{table}

\subsubsection{Code Relation Analysis}
To analyze the impact of the relations in CTGs, we used different variants of CTGs: $\mathcal{R} = \{\textit{structure}\}$, $\mathcal{R} = \{\textit{dependency}\}$, and $\mathcal{R} = \{\textit{structure, dependency}\}$. In this experiment, we used the same setting for JIT-VD model (RGAT model having two layers, and \textit{sum} readout function) for all the variants, and applied in the \dev setting.

As seen in Table~\ref{tab:view_analysis}, \tool performs quite stably when $\mathcal{R} = \{\textit{structure}\}$ and $\mathcal{R} = \{\textit{dependency}\}$, i.e., \textit{F1 = 0.69} in both cases. When combining both relations in CTGs, \tool detects the dangerous commits significantly more precisely while its \textit{recall} remains stable. It is reasonable because considering both the structure relation and the dependency relation in CTGs gains the information that is necessary for the model to determine if a commit is suspiciously dangerous. As a result, both the JIT-VD \textit{F1} and \textit{accuracy} increase when $\mathcal{R} = \{\textit{structure, dependency}\}$. 

Since several existing studies consider the execution flow (i.e., control flow) relation between program elements in detecting vulnerabilities~\cite{gascon2013structural, sparks2007automated}, we additionally consider $\mathcal{R} = \{\textit{structure, dependency, exec. flow}\}$ in CTGs to evaluate the expandability of our proposed code change representation. We found that additionally considering the execution relation in CTGs, $\mathcal{R} = \{\textit{structure, dependency, exec. flow}\}$, decreases the performance of \tool by 1.4\% in \textit{F1}. As expected, in this case, the \textit{precision} is 4.8\% better than that of \tool with $\mathcal{R} = \{\textit{structure, dependency}\}$. Nevertheless, the changes in the execution flow (i.e., control flow), that are required for determining dangerous commits, could be covered by the changes in the dependency relation via control dependencies. This redundancy hurts \tool's recall 6.1\%. Moreover, this conditional consideration increases the complexity of CTGs and the cost for training and testing the JIT-VD model.
Thus, in this work, we use $\mathcal{R} = \{\textit{structure, dependency}\}$ in CTGs for the best performance and efficiency of \tool.

\begin{table}[]
\centering
\caption{Impact of the code relations on \tool's performance}
\label{tab:view_analysis}
\begin{tabular}{@{}l|rrr|r@{}}
\toprule
                        & \textit{Precision} & \textit{Recall}   & \textit{F1}   & \textit{Accuracy} \\ \midrule
Struct.                 & 0.72	             & 0.66	             & 0.69	         & 0.82               \\
Depend.                 & 0.70	             & 0.69	             & 0.69	         & 0.81               \\
Struct.+Depend.         & 0.82	             & 0.66	             & 0.73	         & 0.85	             \\
\bottomrule
\end{tabular}
\end{table}

\begin{gtheorem}
\textbf{Answer for RQ3}: Besides changed code in CTGs, the related unchanged code significantly contributes to the effectiveness of \tool in detecting dangerous commits. Additionally, incorporating \textit{structure} and \textit{dependency} relations in CTGs enhances the JIT-VD performance of \tool. 
\end{gtheorem}

\subsection{Sensitivity Analysis (RQ4)}

\subsubsection{Impact of Training Size}

\begin{figure}
    \centering
    \includegraphics[width=0.9\columnwidth]{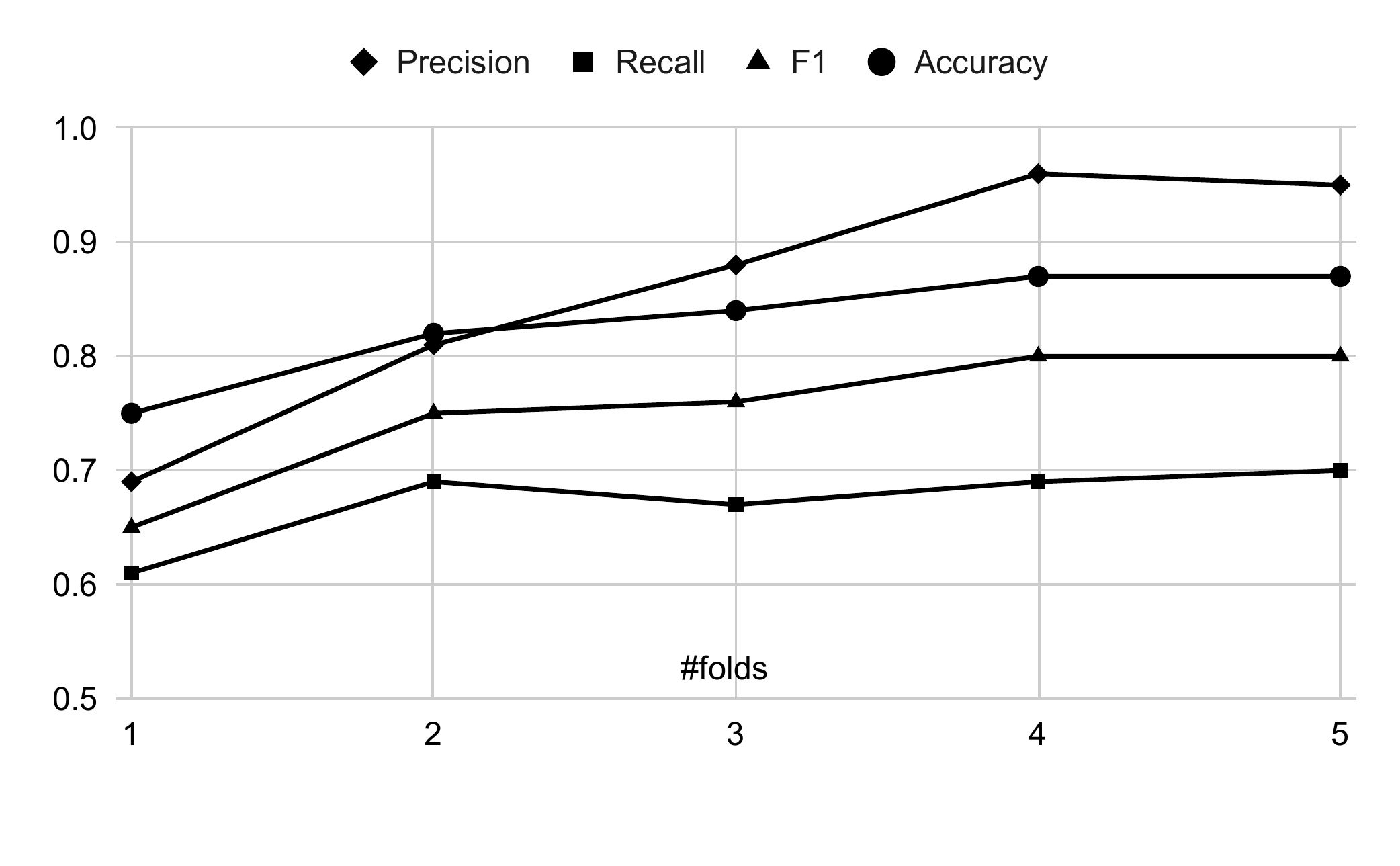}
    \caption{Impact of training data size}
    \label{fig:impact_training_size}
\end{figure}

To measure the impact of training data size on \tool's performance, we use the \cross setting, in which the training set is randomly separated into 5 folds.  We gradually  increased the sizes of the training dataset by adding one fold at a time until all the 5 folds are added for training. The results of this experiment are shown in Fig.~\ref{fig:impact_training_size}.
As expected, the performance of \tool is improved when expanding the training dataset. Especially, the precision significantly grows by 38\% when increasing the training size from 1 fold to 5 folds. The reason is that with larger training datasets, \tool has observed more and then performs better. At the same time, training with a larger dataset costs more time. Particularly, the training time of \tool with 5 folds is six times more than the training time with only 1 fold.

\subsubsection{Impact of Change Complexity}
%
%
%

\begin{figure}
    \centering
    \includegraphics[width=0.9\columnwidth]{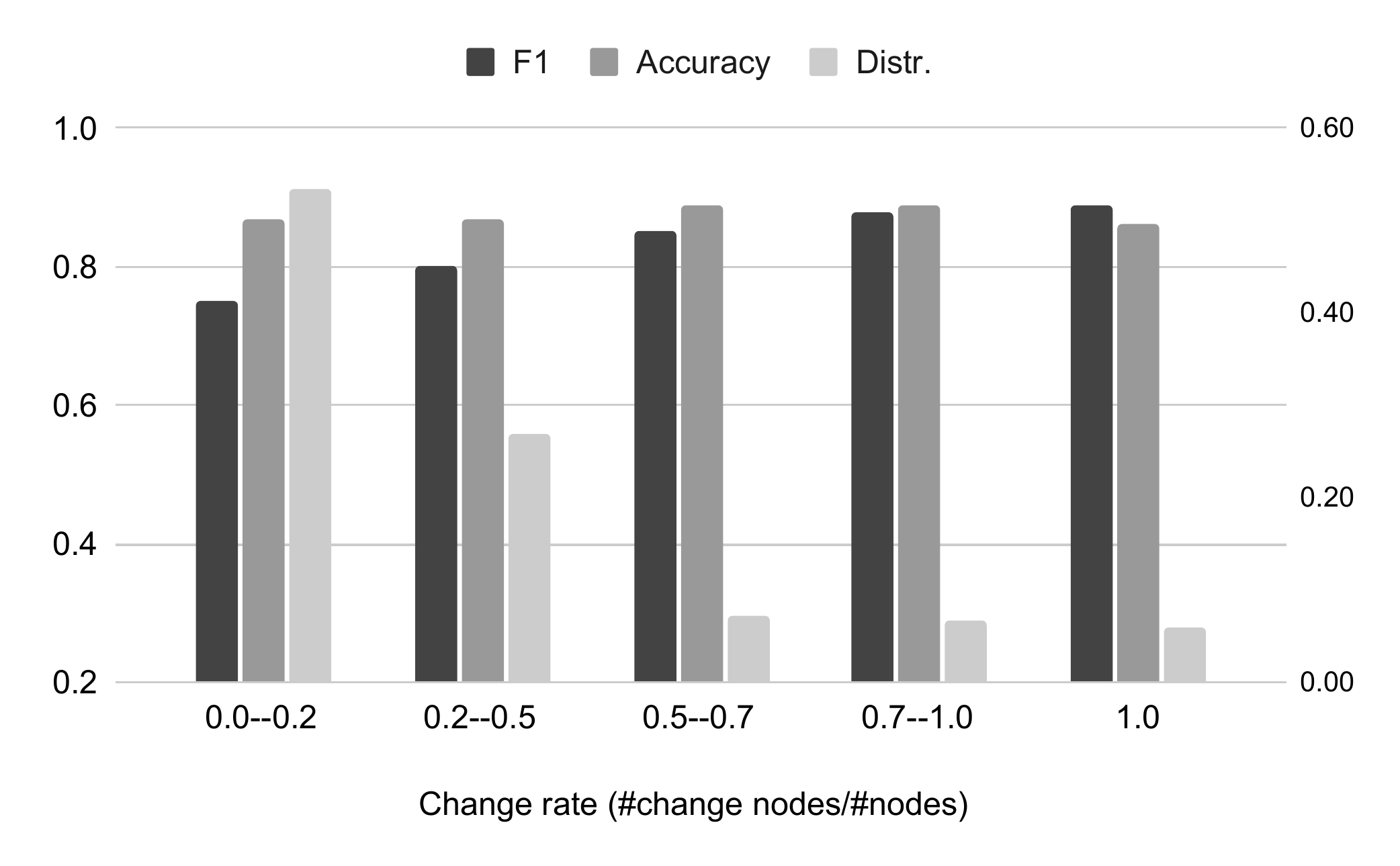}
    \caption{Impact of change complexity (left axis: \textit{F1} \& \textit{accuracy}; right axis: Distr.)}
    \label{fig:impact_change_size}
\end{figure}

In this experiment, we investigate the sensitivity of \tool's performance on the input complexity in change rates which are measured by the ratio of the number of changed nodes over the total number of nodes of each CTG (Fig.~\ref{fig:impact_change_size}). As seen, there are much fewer commits (lower distribution - Distr.) with higher complexity levels. Meanwhile, the performance of \tool in both \textit{F1} and \textit{accuracy} is quite stable when handling commits in different change rates. Specifically, the \textit{F1} of \tool gracefully grows from 75\% to 89\%, and the \textit{accuracy} is in the range of 86\% to 89\%, when increasing the change rate.


\begin{gtheorem}
\textbf{Answer for RQ4}: The JIT-VD performance of \tool improves when \tool is trained on a larger  dataset. The performance of \tool in both \textit{F1} and \textit{accuracy} is stable when classifying commits with different change complexity levels.
\end{gtheorem}

\subsection{Time complexity (RQ5)}
In this work, all our experiments were run on a server running Ubuntu 18.04 with an NVIDIA Tesla P100 GPU. 
On average, \tool took less than a second to construct a CTG.
In \tool, training the JIT-VD model took about 5 hours for 50 epochs. Additionally, \tool with FastRGCN spent 0.75 seconds to classify whether a commit is suspicious or not. 

As expected, the more complicated JIT-VD model, the longer the training time. For a fold of data, 1-layer RGAT model took about 40 minutes, while the figures for 3-layer RGAT model and 5-layer RGAT models are 1.5 times and 1.8 times larger. Also, the larger the training dataset, the longer the training time. Specifically, the training time of five folds of data is six times more than that of only one fold. The classification time is not quite different among the variants of \tool. For example, with FastRGCN, the average classification time is 0.75 seconds, while \tool with RGCN took 1.42 seconds to classify a commit.

\subsection{Threats to Validity}
The main threats to the validity of our work consist of internal, construct, and external threats.

\textbf{Threats to internal validity} include the influence of the method used to construct relational code graph and identify the relation between program entities. To reduce this threat, we use Joern~\cite{joern} code analyzer, which is widely used in existing studies~\cite{velvet,vuldeeppeaker,ivdetect,linevd}. Another threat mainly lies in the correctness of the implementation of our approach. To reduce such threat, we carefully reviewed our code and made it public~\cite{CodeJIT} so that other researchers can double-check and reproduce our experiments.

\textbf{Threats to construct validity} relate to the suitability of our evaluation procedure. We used \textit{precision}, \textit{recall}, \textit{F1}, and classification \textit{accuracy}. They are the widely-used evaluation measures for \jit defect detection~\cite{jit-fine,jitline,lapredict} and \jit vulnerability detection~\cite{vccfinder,vuldigger}. 
Besides, a threat may come from the adaptation of the baseline approaches. To mitigate this threat, we directly obtain the original source code from the GitHub repositories of the studied techniques. Also, we use the same hyperparameters as in the original papers~\cite{jit-fine,jitline,lapredict,vccfinder,deepjit}. 

\textbf{Threats to external validity} mainly lie in the selection of graph neural network models used in our experiments. 
To mitigate this threat, we select the representative models which are well-known for NLP and SE tasks~\cite{rgcn,rgat,fastgcn}. 
In addition, the construction of our dataset is another threat which is also suffered by related studies~\cite{lapredict, vccfinder, are_we_there}. 
Specifically, the commits triggering/contributing to vulnerabilities are considered as dangerous ones.
%
%
To reduce the impact of this threat, we focus on the vulnerability triggering commits as the dangerous commits instead of vulnerability contributing commits. Also, to precisely blame vulnerable statements during collecting dangerous commits, we consider program dependencies instead of surrounding statements as mentioned in Sec.~\ref{sec:data}. 
Another threat is considering the vulnerability-fixing commits as safe ones because these commits could contribute to vulnerabilities. To reduce this threat, only the fixing commits which are not VCCs are considered as safe.
Moreover, our experiments are conducted on only the code changes of C/C++ projects. Thus, the results could not be generalized for other programming languages. In our future work, we plan to conduct more experiments to validate our results to other languages. 



\section{Related Work}
\tool relates to the work on \textbf{just-in-time vulnerability detection}~\cite{vccfinder, vuldigger}. 
VCCFinder~\cite{vccfinder} and VulDigger~\cite{vuldigger} are the classification models combining code-metrics, commit messages, and expert features for JIT-VD.
Meanwhile, the work on \textbf{just-in-time bug detection}~\cite{jit-fine,jitline,lapredict,deepjit} could also be applied to detect vulnerabilities at the commit-level. DeepJIT~\cite{deepjit} automatically extracts features from commit messages and changed code and uses them to identify defects. Pornprasit~\etal propose JITLine, a simple but effective just-in-time defect prediction approach. JITLine utilizes the expert features and token features using bag-of-words from commit messages and changed code to build a defect prediction model with random forest classifier. 
LAPredict~\cite{lapredict} is a defect prediction model by leveraging the information of "lines of code added" expert feature with the traditional logistic regression classifier. 
Recently, Ni~\etal introduced JITFine~\cite{jit-fine} combining the expert features and the semantic features which are extracted by CodeBERT~\cite{codebert} from changed code and commit messages. 
Different from all prior studies in \jit bug/vulnerability detection, our work presents the first study centralizing the role of code change semantics in detecting vulnerabilities at the commit level. In \tool, the code change meaning is the only deciding factor in examining the suspiciousness of commits. Our results experimentally show that our strategy is more appropriate for JIT-VD.


\textbf{Vulnerability detection} is also a critical part to be discussed. Various methods have been proposed to determine if a code component (component, file, function/method, or statement/line) is vulnerable. 
Recently, several deep-learning based approaches have been introduced~\cite{survey_papers, poster, vuldeeppeaker, vulsniper, devign, are_we_there,mvd,issta_22}. 
VulDeePecker~\cite{vuldeeppeaker} and SySeVR~\cite{sysevr} introduce tools to detect \textit{slice-level} vulnerabilities, which are more fine-grained.
IVDetect~\cite{ivdetect}, which is a graph-based neural network model, is proposed to detect vulnerabilities at the function level and use a model interpreter to identify vulnerable statements in the detected suspicious functions. LineVul~\cite{linevul} and LineVD~\cite{linevd} apply CodeBERT in their own way and have been shown that they are more effective than IVDetect in detecting vulnerable functions and lines/statements. VelVet~\cite{velvet} builds graph-based models to detect vulnerable statements.
Those approaches focus on detecting vulnerabilities at the release-time, while \tool is a \jit vulnerability detection. \tool and the release-time approaches could complement to support developer in ensuring software quality in the development process.  



Several \textbf{learning-based approaches} have been proposed for specific SE tasks including code suggestion~\cite{icse20, naturalness, allamanis2016convolutional}, program synthesis~\cite{amodio2017neural,gvero2015synthesizing}, pull request description generation~\cite{hu2018deep,liu2019automatic}, code summarization~\cite{iyer2016summarizing,mastropaolo2021studying,wan2018improving}, code clones~\cite{li2017cclearner}, fuzz testing\cite{godefroid2017learn}, bug detection~\cite{oppsla19}, and program repair~\cite{jiang2021cure,ding2020patching}.

\section{Conclusion}
In this paper, we propose \tool, a novel code-centric approach for just-in-time vulnerability detection (JIT-VD). 
Our key idea is that \textit{the meaning of the changes in source code caused by a commit is the direct and deciding factor for assessing the commit's riskiness.}
We design a novel graph-based multi-view code change representation of the changed code of commits in relation to the unchanged code. Particularly, a graph-based JIT-VD model is developed to capture the patterns of dangerous/safe (benign) commits and detect vulnerabilities at the commit level. 
On a large public dataset of dangerous and safe commits, our results show that \tool significantly improves the state-of-the-art JIT-VD methods by up to 66\% in \textit{Recall}, 136\% in \textit{Precision}, and 68\% in \textit{F1}. 
Moreover, \tool correctly classifies nearly 9/10 of dangerous/safe  commits. 

\bibliographystyle{IEEEtran}

\bibliography{main}
\balance
\end{document}